# Towards Modeling Energy Consumption of Xeon Phi

**Gary Lawson   Masha Sosonkina   Yuzhong Shen**
Old Dominion University
1300 Engineering and Computational Sciences Building
{glaws003, msosonki, yshen} @ odu.edu

**ABSTRACT**
In the push for exascale computing, energy efficiency is of utmost concern. System architectures often adopt accelerators to hasten application execution at the cost of power. The Intel Xeon Phi co-processor is unique accelerator that offers application designers high degrees of parallelism, energy-efficient cores, and various execution modes. To explore the vast number of available configurations, a model must be developed to predict execution time, power, and energy for the CPU and Xeon Phi. An experimentation method has been developed which measures power for the CPU and Xeon Phi separately, as well as total system power. Execution time and performance are also captured for two experiments conducted in this work. The experiments, frequency scaling and strong scaling, will help validate the adopted model and assist in the development of a model which defines the host and Xeon Phi. The proxy applications investigated, representative of large-scale real-world applications, are Co-Design Molecular Dynamics (CoMD) and Livermore Unstructured Lagrangian Explicit Shock Hydrodynamics (LULESH). The frequency experiment discussed in this work is used to determine the time on-chip and off-chip to measure the compute- or latency-boundedness of the application. Energy savings were not obtained in symmetric mode for either application.

**INTRODUCTION**
Energy consumption is of the utmost concern with respect to exascale computing. Accelerators may be used to quicken execution at the cost of power, and this tradeoff may lead to energy savings. The Intel MIC (many-integrated core) architecture is at the heart of the Intel Xeon Phi co-processor. The co-processor is typically composed of many low-frequency, energy-efficient, in-order cores. The high parallelism and energy-efficient design make the device an appealing option for accelerated computing; hence the investigation of the Intel Xeon Phi in this work. The authors aim to develop a method for predicting the best configuration for any application, with any set of optimizations, which utilizes the Xeon Phi co-processor. To make such predictions, a model must be developed which explains the execution time, power, and energy consumption of an application given a computing platform. This work lays the groundwork for developing such a model and method.

The purpose of this work is to develop a model which may be used to predict energy for a computing platform containing Xeon Phi. This paper will present the base CPU time and power models and extend the power model to include the measured Xeon Phi power. This paper will also present a new experimentation method which measures total system power, CPU energy, and Xeon Phi power during execution. The benefits of the change in measurement schemes will be discussed.

**Background**
The Intel Xeon Phi is composed of 50+ cores at approximately 1 GHz. Each core is capable of concurrently processing four hardware threads [1]. Threads may be mapped to cores through the *affinity* environment variable [2]. Each core is also equipped with a 512-bit vector processing unit capable of processing up to 8 or 16 operations per second depending on data precision, double- and single-precision respectively.

The Xeon Phi device supports various execution modes: native, offload, and symmetric. Native mode executes directly on the device. Offload mode executes code sections declared using compiler directives and requires manipulation of the code. Symmetric mode executes on both the Xeon Phi and host. Execution begins on the host and the Xeon Phi is treated as an additional node; MPI tasks are assigned to each device, and all communications occur over the PCIe bus. Symmetric mode execution is explored in this work.

To reduce energy consumption by the CPU, a dynamic voltage and frequency scaling (DVFS) technique is commonly used at the application runtime (see, e.g., [3]). The current generation of Intel processors provides various P-states for frequency scaling. Frequency may be specified by manipulating the model-specific registers (MSR). P-states are defined as $f_i, \dots, f_n$ where $f_i > f_j$, $0 < i < j \leq n$. For the older generations of the Intel Xeon Phi, DVFS is not software-accessible. Therefore, methods to improve the overall performance must be explored to minimize total energy.

Proxy applications are miniature codes which are built to be representative of the workload characteristics of a full production code. The application code is generally provided in C/C++/Fortran with hybrid MPI and OpenMP implementations. Those proxy applications which have been adopted by the community may also include other paradigms, such as CUDA implementations for GPU computing. The proxy applications used in this work are Co-

Design Molecular Dynamics (CoMD) [4], [5] and Livermore Unstructured Lagrangian Explicit Shock Hydrodynamics (LULESH) [6], [7]. For each application, two executables are compiled: one for the host and the other for the Xeon Phi. Both executables are required to run the application in symmetric execution mode. The base hybrid MPI and OpenMP implementation was used for each application.

**Related Work**

Measuring the power draw of the Xeon Phi may be accomplished by a few methods/tools. The first tool investigated by the authors was MICSMC [8], however the tool increases the power draw of the device during measurement [9]. Power may also be measured by reading the "/sys/class/micras/power" file which is updated every 50ms [10], [11]. Finally, power may be obtained via the MICAccessSDK API provided by Intel [11], [12].

Modeling CPU energy based upon host frequency was recently investigated [13] and serves as the base model for this work. The authors perform linear regression to determine time on- and off-chip, as well as static and dynamic power for each application. These parameters are then used to determine the optimal frequency.

Power and/or performance evaluations of the Intel Xeon Phi include [11], [14], [15], [16], [18], and [20]. In [11], the authors investigate the power and performance of auto-tuned applications, including a modified version of LULESH (v1.0), to determine the optimal energy on computing platforms which include the Intel Xeon Phi. However, LULESH was executed for the host CPU, and not for the Intel Xeon Phi. In this work, LULESH (v2.0) is compiled and executed for CPU and CPU+MIC configurations.

Modeling of power and performance of architectures including the Xeon Phi was conducted in [14] where extension of the roofline model was applied. The work presented in [15] measures power and performance for the Rodina and SHOC benchmarks on multi-core CPU, Xeon Phi, and GPU architectures. A performance evaluation of An instruction-level energy model for the Intel Xeon Phi was proposed in [17]. A high performance MPI library for clusters featuring the Intel Xeon Phi and Infiniband was proposed in [18] which is based on the (symmetric communications interface) SCIF API [19]. An investigation of performance on molecular dynamics applications for the Intel Xeon Phi has been conducted in [20].

The remainder of the paper is organized into the following sections. The *Energy Model* is presented which includes the Xeon Phi power measurements and presents the parameters to be captured in the experiments conducted. The *Experimental Methodology* section discusses the various components which make up the experimental method. The *Results* section presents the energy and power measured for each application and each configuration experimented. Finally, the *Conclusion* concludes the document and presents the future research direction.

**EXISTING ENERGY MODEL**

This section discusses the CPU energy model adopted by the authors which serves as the base model for this work, and future work. Adaptation to model the Xeon Phi execution time and power is the goal for future research. Such a model would provide valuable information toward determining a best configuration for a given application without having to execute every configuration variation imaginable. The initial energy model will describe the response of the host CPU with respect to frequency change for a given application. The host CPU energy model is composed of a time and power model; the product of each equates to energy.

**Time Model**

The CPU time model, as described in [13], [21] is:

$$T_f = (t_{on} f_{max})/f_i + t_{off}. \quad (1)$$

It provides valuable insights into the workload and usage of the host CPU by separating the time on-chip ($t_{on}$) from the time off-chip ($t_{off}$) during execution. The maximum frequency ($f_{max}$) and current frequency ($f_i$) are used to determine the slope. The current frequency may not be equal to the maximum frequency for this linear relationship. The reported execution times from the frequency scaling experiment are used to determine $t_{on}$ and $t_{off}$ using linear regression because time on-chip scales linearly with the host frequency.

The Xeon Phi, with respect to the host, simply contributes to time off-chip because time spent waiting for the co-processor contributes to latency experienced by the host. If the Xeon Phi does not impact the host during execution, $t_{on}$ and $t_{off}$ should be similar to the host execution. The Xeon Phi will impact the host however, because the executed applications have not been adapted to accommodate the Xeon Phi in this work. Load balance between MPI tasks is expected to be an issue; load imbalance would cause the Xeon Phi or host to wait on the other. Calculating $t_{on}$ and $t_{off}$ for the results of each execution will provide a measurable metric which may be used to compare the different executions.

**Power Model**

Until now, power has been measured and the recordings used directly to calculate energy consumed. Time was read from the application output file, and power was measured using the Wattsup meter [22]. Because the sampling rate for Wattsup is one second, the values were summed over the execution interval. However, this is not an effect method for acquiring energy because it does not provide a means of estimating power for future configurations.

The CPU power model from [13] has been adopted in this work as the base power model, as defined below.

$$P_f = P_S + knf^3 \quad (2)$$

$P_S$ is the static power draw, a constant power which is drawn independent of CPU execution. The dynamic power draw is impacted by CPU frequency and core usage. It consists of frequency $f$ cubed, the number of cores $n$, and the model constant $k$.

To include the Xeon Phi, $P_{MIC}$ is simply the sum of the average power of each Xeon Phi device used during execution. Xeon Phi power draw is independent of the host CPU frequency, and therefore the measured result may be used directly until a model has been developed to estimate power for the Xeon Phi.

**Energy Model**
Energy is thusly defined as:
$$E = E_{CPU} + E_{MIC}, \text{ where} \quad (3)$$
$$E_{CPU} = T_f \times P_f \quad \& \quad E_{MIC} = T_f \times P_{MIC}.$$

The Xeon Phi energy model uses the time defined by the model $T_f$, total execution time with respect to host frequency that includes any Xeon Phi induced latency, and $P_{MIC}$.

For CPU energy, calculating the product between the time model and power model yields the following fourth-order equation:
$$af^4 + bf^3 + cf^2 + df + e = 0, \text{ where} \quad (4)$$
$$a = 1 \quad b = \frac{2t_{on}f_{max}}{3t_{off}} \quad c = d = 0 \quad e = -\frac{t_{on}f_{max}P_S}{3knt_{off}}.$$

To obtain the optimal frequency from the closed form expression, energy must be differentiated with respect to host frequency. For the case of $t_{off} = 0$, the CPU energy model simplifies to:
$$E_{CPU} = \frac{t_{on}f_{max}P_S}{f} + t_{on}f_{max}knf^2. \quad (5)$$

And therefore optimal frequency is:
$$f_{opt} = \sqrt[3]{P_S/2kn} \quad (6)$$

For the case of $t_{off} > 0$, optimal frequency may be obtained via the solution explained thoroughly in [13].

To validate the model, a frequency scaling experiment has been conducted to perform measurements on power, time, and performance. The time and power model parameters $t_{on}$, $t_{off}$, $k$, and $P_S$ will all be calculated, as well as the optimal host frequency.

## EXPERIMENTATION METHODOLOGY

This section presents the experimentation method used to conduct the frequency and problem size scaling experiments discussed in the following section. The purpose for these experiments is to gain insights into the response of an application when subjected to changes in the execution configuration. For the time model, $t_{on}$ and $t_{off}$ are determined by the frequency scaling experiment; for the power model, $k$ and $P_S$ are determined. The problem size scaling experiment is used to investigate the reaction of an application to varying the workload per task (i.e. strong scaling). Energy, power, execution time, and performance are measured during execution.

**Measuring Power**
This section will present the methods for measuring power from each source: *Wattsup*, *Cpu*, and *Mic*.

**Wattsup:** The Wattsup meters are connected to the system via USB connection. The data is sampled at a rate of 1Hz; this resolution is course, however most executions are sufficiently long (> 30 seconds) and energy is no longer measured using this data. The Wattsup power data provides total system power, including cooling, hard drives, network devices, etc. Reducing the power draw of these devices directly is not within the scope of this work. Hence, Wattsup power is used as a reference to measure overhead as a result of performing power measurements.

**CPU:** The sandy-bridge processors used in this work support power measurement via the RAPL (Running Average Power Limit) interface. In Linux, the RAPL energy readings are available via the Model Specific Registers (MSR). RAPL provides energy measurements made for the entire package, and for the core and DRAM components. Depending on the model processor, uncore components may also be measured. In this work, CPU package power is used. Energy is reported directly in joules and is sampled at a rate of 50ms [23]. Average power is calculated for each sample window.

**Xeon Phi:** The Xeon Phi co-processor provides power measurements directly in the form of a file available in the system directory. The file (/sys/class/micras/power) provides a wealth of information, but most importantly are the power draw readings for each power connector. For current models of the Xeon Phi, three connectors supply the device: the PCIe, 2x3 pin, and 2x4 pin connectors. The sum of the power measured for each connector equates to the total power of the device [12]. The file may be sampled at a rate of 50ms.

**Experimentation Method**
An experiment is composed of a collection of individual tests, each executed with a specific configuration. For each test, there exists a list of parameters which may be varied as shown in Table 1. For an experiment, a subset of the parameters are varied; each parameter will be specified when detailing each experiment.

| Experiment Parameters | |
|---|---|
| Problem Size | # Nodes |
| # MPI Tasks | # Xeon Phi |
| MIC Affinity | MIC Execution Mode |
| Host Frequency | Host DVFS Frequency |
| # Host OMP Threads | # MIC OMP Threads |

Table 1. List of experiment parameters

For each test, the configuration parameters are established before the execution and measurements begin. Following, the measurements are started in two tiers. First, the Wattsup measurement begins; this measurement reads both meters connected to the system and stores the output in a file. After executing the Wattsup measurement, 15 seconds of idle time is provided to begin collecting baseline power measurements. The CPU and MIC power readers are then executed. For the CPU, a thread is spawned on the host to read the MSR for the RAPL energy output. For the MIC, a thread is spawned on each Xeon Phi which simply reads the power file and outputs the data, with timing information, to a CSV (comma separated value) format for parsing. Another 15 seconds are provided to collect pre-execution measurements. Wattsup measurements made during this time will report power draw differences when performing component specific measurements.

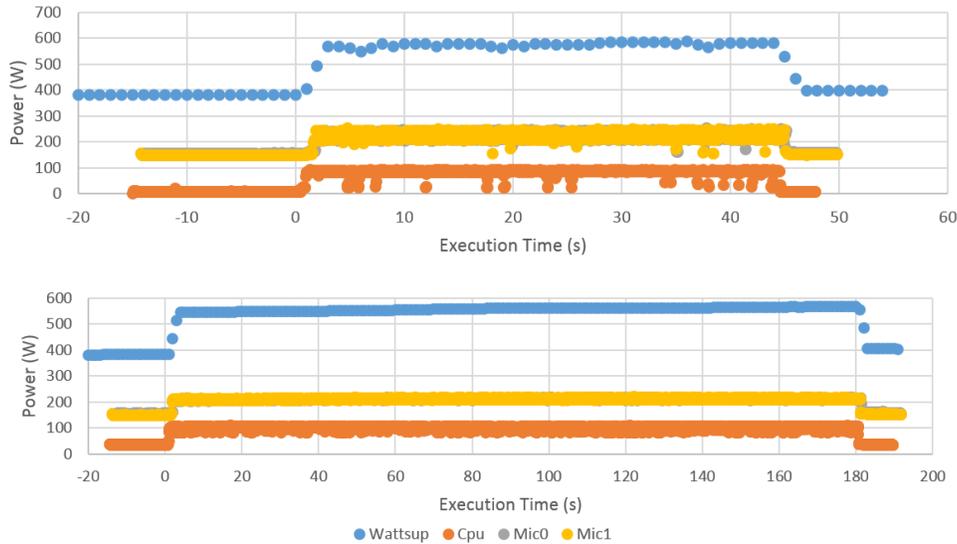

**Figure 1. Power profiles for the execution of CoMD (top) and Lulesh (bottom) given two Xeon Phi.**

Following the second idle period, the application is executed as per for the configuration specifications. Measurements persist throughout execution; however the Wattsup power meter often has difficulties providing a stable stream of power readings. Hence, energy is now based upon the CPU and MIC power reads. Upon execution completion, the system will remain idle for 20 seconds and then the measurements are terminated. Following, the host remains idle for 60 seconds to allow the system to cool, and each component to return to a stable power draw.

For MIC, the power output files must be copied back from the device and stored on the host. The output files may require up to 50 MB for executions lasting longer than 30 minutes, but generally are only a few MB in size and do not waste the limited MIC resources. Should this become a problem in the future, SCIF may be employed to copy back readings to the host in real-time.

Figure 1 presents the power profiles for CoMD and LULESH when executed with two Xeon Phi in symmetric mode. The profiles were taken from the samples collected during the frequency scaling experiment to be discussed. When observing the Wattsup power readings, power draw is not increased by reading CPU or MIC power. Previously, MICSMC was used to obtain power measurements from the co-processor; however the tool incurred a high power draw penalty. However, the current measurement method does not incur a power draw penalty. Additionally, it does not incur a performance penalty either because executions performed in this work only occupy 236 out of 240 possible threads. One core remains available for the micro-OS on the device. The power measurement thread will occupy one of the four available threads on the remaining core, thus avoiding resource sharing with the executed application being tested.

**RESULTS**

This section presents the results for the frequency and problem size scaling experiments. The frequency scaling experiment will present energy and power for each application measured during execution. The same will be presented for the problem size experiment. Finally the calculated model parameters from the frequency experiment will also be presented.

The tests conducted in each experiment have been grouped into four main configurations: Host 1, Host 2/8, MIC 1, and MIC 2. Host-only executions are denoted with *Host* and those including a Xeon Phi are denoted with *MIC*. For CoMD, one to three MPI tasks are required for execution; the host-only executions explore one and two MPI task implementations whereas the Xeon Phi executions require an additional MPI task for each MIC device (and one for the host because the execution mode is symmetric).

For LULESH, one and eight MPI tasks may be used due to requirements imposed by the application. The host-only executions explore the use of one and eight MPI tasks on the platform. Since the MIC executions require as few as two MPI tasks, eight MPI tasks must be allocated to run the application. One MPI task is assigned to each MIC device and all remaining tasks are assigned to the host. Multiple MPI tasks may be assigned to the Xeon Phi devices, however it is inefficient due to slow communications between Xeon Phi devices and the host or other Xeon Phi. Although a formal experiment has not been conducted to explore the MPI task distribution and documented, MPI task distribution has been investigated by the authors. The results showed MPI communications over the PCI-E bus with the default Intel MPI library slow execution exponentially with the number of devices communicating with one another (for LULESH especially). For this reason, one MPI task is assigned per device.

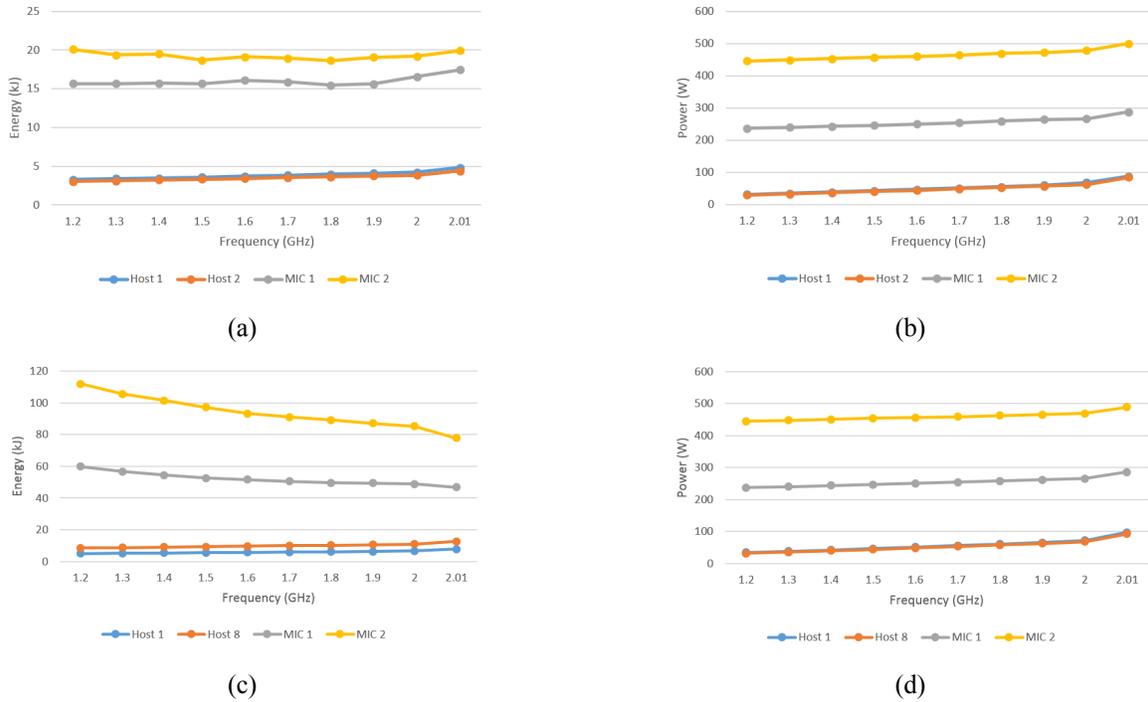

**Figure 2. Energy (a, c) and average power (b, d) for CoMD (a, b) and LULESH (c, d) during the frequency scaling experiment.**

For each experiment, affinity is set to *compact*, problem size is set to $60^3$ elements (LULESH) and $4*60^3$ atoms (CoMD), 236 threads are specified on the Xeon Phi, 16 threads are specified on the host, and the host frequency is set to the maximum, 2.01 GHz. DVFS is not implemented in this work.

**Test Platform**
Experiments were performed on the single-node "Borges" computing platform at Old Dominion University. The Borges system contains two Intel Xeon E5-2650 8-core processors with HyperThreading technology, each core runs at 2GHz (2.8GHz Turbo), has 20MB of L3 cache, and 64GB total of RAM. The Intel Xeon E5-2650 processor supports 10 P-states and the frequencies range from 1.2 GHz to 2.01 GHz. A 100MHz stepping is used for all the states except the last one, where only a 1MHz stepping is used. "Borges" contains two Intel Xeon Phi co-processors 5110P, each connected to the host via a PCIe, and have 30MB of the L2 cache, 32KB of the L1 data and instruction caches, 8GB DRAM (GDDR5), and 60 cores running at 1.05GHz. Each core contains a wide 512-bit single instruction multiple data (SIMD) vector processing unit, which implements the fused multiply-add operation; 8 double-precision or 16 single-precision data elements may be operated upon in a single execution phase. Two Wattsup power meters supply the platform with power.

**Frequency Scaling**
Figure 2 presents the energy and average power for CoMD, and LULESH for the frequency scaling experiment. Host frequency is varied from the minimum to maximum and performs several tests at each frequency level. The results for each of the four configurations are presented within each plot. References to CoMD in this section will refer to Figure 2 (a, b) and the energy and average power plots thereof. References to LULESH in this sub-section refer to Figure 2 (c, d).

Energy consumed by CoMD for the various frequency states reached a minimum at 1.8 GHz for the Xeon Phi executions; the host executions obtained a minimum energy at 1.2 GHz. LULESH suffers from the large number of MPI tasks required as shown in the energy plot. The minimum energy is obtained when frequency is max; this shows that energy is bounded explicitly by execution time for Xeon Phi executions. Host executions follow a pattern similar to CoMD in that the minimum frequency presented the lowest energy consumed. The host shows a considerable increase in energy consumption when increasing from one to eight MPI tasks because the host's computational resources are saturated. Average power for CoMD and LULESH increases linearly with frequency; both plots show Xeon Phi power increases the average power draw by approximately 200W per Xeon Phi. Because MIC frequency is independent of the host frequency, this power draw is constant as host frequency increases.

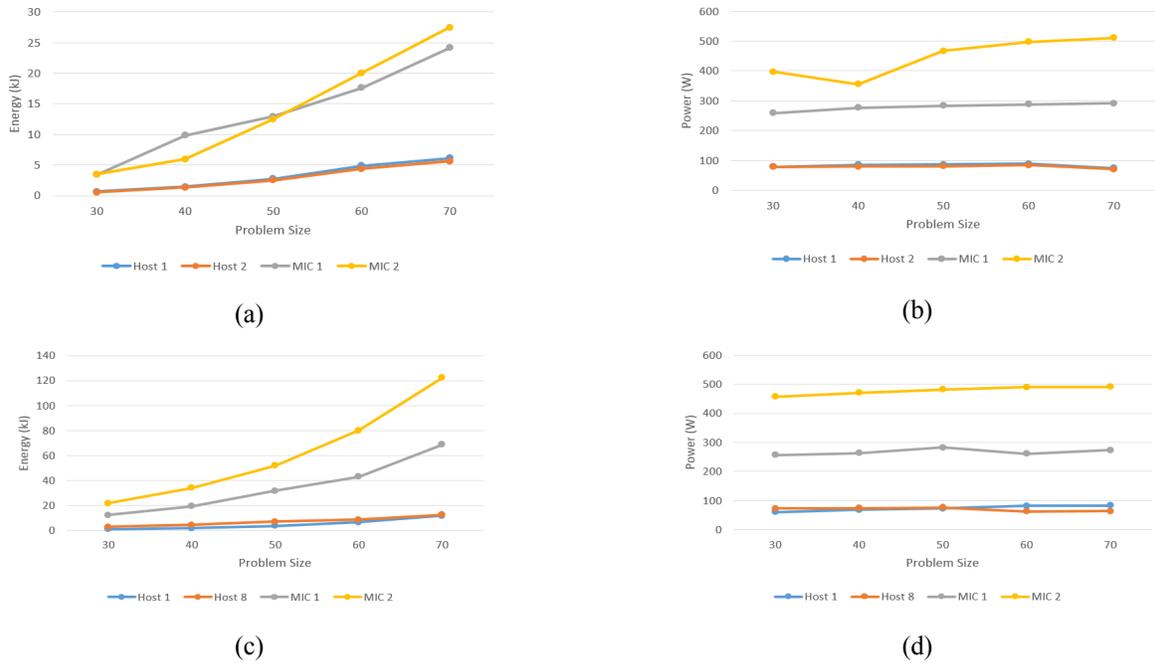

**Figure 3. Energy (a, c) and average power (b, d) for CoMD (a, b) and LULESH (c, d) during the problem size scaling experiment.**

| | CoMD | | | | | | LULESH | | | | | |
|---|---|---|---|---|---|---|---|---|---|---|---|---|
| Config | $t_{on}$ | $t_{off}$ | $t_{off}/t_{on}$ | $k$ | $P_S$ | $R^2$ | Config | $t_{on}$ | $t_{off}$ | $t_{off}/t_{on}$ | $k$ | $P_S$ | $R^2$ |
| Host 1 | 64.19 | 0.61 | 0.01 | 0.34 | 23.09 | 0.999 | Host 1 | 77.42 | 11.02 | 0.14 | 0.38 | 26.07 | 0.998 |
| Host 2 | 61.14 | 0.00 | 0.00 | 0.32 | 21.99 | 0.999 | Host 8 | 150.93 | 8.81 | 0.06 | 0.36 | 24.67 | 0.999 |
| MIC 1 | 13.21 | 55.41 | 4.19 | 0.29 | 22.94 | 0.491 | MIC 1 | 118.35 | 86.41 | 0.73 | 0.29 | 22.36 | 0.991 |
| MIC 2 | 9.15 | 33.78 | 3.69 | 0.31 | 22.97 | 0.735 | MIC 2 | 115.89 | 84.00 | 0.72 | 0.26 | 20.10 | 0.993 |

**Table 2. List of model parameters and calculated values**

**Strong Scaling**

Figure 3 presents the energy and average power for CoMD, and LULESH for the problem size scaling experiment. Problem size is varied from 30 to 70 to show how the system interacts with varying workloads. The results for each of the four configurations are presented individually within each plot. References to CoMD refer to Figure 3 (a, b), LULESH to Figure 3 (c, d) in this sub-section.

Energy consumption for the host-only execution of each application remains fairly constant, with slight increases or decreases for select problem sizes. For the MIC executions, the minimum energy consumption is obtained with the lowest problem size (this result is consistent for both applications). MIC executions of CoMD consume as low as 4x energy compared to the host, and LULESH consumes as low as 10x energy. The cause for the dramatic increase in energy consumption with problem size is the quick saturation of MIC resources due to inefficient vectorization usage and communications between tasks and devices over the PCI-E bus.

In CoMD, the decrease in energy consumption and power for problem size 40 is the result of the execution exiting improperly. At some point between the 20[th] and 30[th] iteration, the application exits. It is possible the application should not be run with this problem size using three MPI tasks.

Average power for both CoMD and LULESH show a slight increase in power draw as problem size increases for MIC executions. Host executions remain fairly constant until problem size 70 (CoMD) and 60 (LULESH) where host power decreases slightly.

**Model Parameters**

By performing linear regression on the frequency scaling results for execution time and CPU power draw, the model parameters presented in Table 2 may be calculated. The ratio, $t_{off}/t_{on}$, may be used to determine how compute- or latency-bounded an application is. Typically, an application is latency bounded if $t_{off}/t_{on}$ is greater than one; a value of one represents the application is evenly bounded by computation and latency causing operations (i.e. memory stalls). From Table 2, both applications are measured to be computationally intensive. However, the MIC executions for CoMD cause the application to become latency-bounded due to the host waiting for the Xeon Phi tasks to complete the current iteration. The MIC execution for LULESH also caused significant idle time for the host; however the idle time does not outweigh the compute time.

In general, this result shows that both applications suffer from load imbalance between the host tasks and MIC tasks; an expect result considering the application code has not been adapted for the Xeon Phi symmetric mode execution. LULESH suffers overwhelmingly from an oversubscription of MPI tasks to accommodate the Xeon Phi in this execution mode. This may be shown by the dramatic increase in compute time between Host 1 and Host 8. The time off-chip decreases only slightly with the additional MPI tasks, but time on-chip increases two-fold; hence the host threads are saturated. LULESH is expected to perform better in a cluster environment, a topic for future work.

The confidence for the parameters calculated for each configuration is presented in Table 2 ($R^2$). For CoMD, the MIC 1 and MIC 2 configurations have low confidence because the Xeon Phi induced latency is not impacted by frequency. The execution time model, equation 1, does not consider the MIC latency and assumes it is included with $t_{off}$. This assumption is incorrect; MIC induced latency should be factored from $t_{off}$. Estimation of execution time should be performed independent for the Xeon Phi and CPU, where total runtime is defined by the larger estimate. This is to be investigated further in the future.

The static power measurement $P_S$, from Table 2, correlates with the CPU plot presented in Figure 1. Idle CPU power draw is approximately 20W. The variations in $P_S$ can be explained by the inclusion of uncore and DRAM power in the CPU power measurement. The package energy counter was read to determine energy consumption for the entire chip. Alternatively, core power and DRAM power could be investigated independently but that is not advantageous toward the work described in this paper. It may be reconsidered in future variations of the model.

## CONCLUSIONS AND FUTURE WORK

This work provides a complete experimentation methodology for obtaining CPU and MIC measurements during application execution. Additionally, an existing CPU execution, power, and energy model have been introduced and investigated for both host and MIC executions. Future implementations involving symmetric mode will provide timing data for host and MIC nodes separately such that MIC and CPU execution times may be modeled independently.

A measurement method for obtaining system, CPU, and MIC power and energy has been provided. Although the sampling rate for the system remains course, it has been used to validate the CPU and MIC measurement methods. For both methods, the act of measuring power did not significantly increase the system power draw. For the Xeon Phi, the device remained idle, but not in a deep idle state. The previous measurement method, MICSMC, incurred a high power draw penalty by activating all device cores. This raised the device from idle to active, resulting in a large overhead.

In this work, only the host CPU was modeled. The measured Xeon Phi power draw is used directly. For now, the Xeon Phi used a constant number of threads; however future research directions will include again varying affinity and thread count to determine whether or not a more optimal solution exists over the current. The current affinity setting is *compact* and thread count is 236 for both applications.

To determine an optimal solution, $t_{on}$, $t_{off}$, and $P_S$ will be used as metrics for each configuration considered. To obtain this information however, a real-time method for estimating $t_{on}$, $t_{off}$, and $P_S$ will need to be created to reduce the number of executions required. Each parameter may be calculated from the same execution as time and power data will be collected. Further, a model will need to be devised which determines the minimum energy based on the calculated metrics.

The energy model will be improved by devising a linear model for the Xeon Phi based on the number of active cores because frequency cannot be varied. This will provide a $k$ and $P_S$ metric for the Xeon Phi, although $t_{on}$ and $t_{off}$ may not be calculated for the device.

## ACKNOWLEDGMENTS
This work was supported in part by Ames Laboratory and Iowa State University under the contract DE-AC02-07CH11358 with the U.S. Department of Energy. This work was also supported in part by the Air Force Office of Scientific Research under the AFOSR award FA9550-12-1-0476, and by the National Science Foundation grants NSF/OCI---0941434, 0904782, 1047772.